\documentclass[conference]{IEEEtran}
\IEEEoverridecommandlockouts
\usepackage{cite}
\usepackage{amsmath,amssymb,amsfonts}
\usepackage{algorithmic}
\usepackage{graphicx}
\usepackage{textcomp}
\usepackage{xcolor}
\usepackage{url}
\usepackage{balance}
\usepackage{multirow}
\newcommand{\frange}{6--12~GHz}   % change to 6--16~GHz if upper bands are recovered
\newcommand{\nbands}{six}         % number of 1-GHz sub-bands analysed
\def\BibTeX{{\rm B\kern-.05em{\sc i\kern-.025em b}\kern-.08em
    T\kern-.1667em\lower.7ex\hbox{E}\kern-.125emX}}
\begin{document}

% \title{FR3 Channel Measurements in Indoor Factory: Wideband SISO Characterization from 7–18 GHz and MIMO Analysis at 7–8.1 GHz\\

% Just a proposal :D
\title{Wideband Channel Measurements in the Lower FR3 Band (6–12 GHz) for Industrial Environments\\

\thanks{This work was supported in part by the German Federal Ministry of
Research, Technology and Space (BMFTR) in the course of the 6GEM+ Transfer Hub 
under grant 16KIS2411 and in part by the Deutsche Forschungsgemeinschaft (DFG, German Research Foundation) under Project–ID 287022738 TRR 196 for Project S03.}
}

\author{\IEEEauthorblockN{Faizan Rauf, Adam Umra, Oways Alsoloh and Aydin Sezgin}
\IEEEauthorblockA{Institute of Digital Communication Systems, Ruhr University Bochum, Germany\\
\{faizan.rauf, adam.umra, oways.alsoloh, aydin.sezgin\}@rub.de}}

\maketitle
\begin{abstract}
The upper mid-band, also known as Frequency Range 3 (FR3), spans 7.125–24.25 GHz and is a key candidate band for sixth-generation (6G) wireless communications. However, channel measurements in industrial environments remain scarce, and the 3GPP indoor-factory (InF) channel model, developed for omnidirectional antennas, has yet to be validated across the FR3 band. This paper presents a wideband channel measurement campaign in an industrial production hall, covering 135 line-of-sight (LOS) positions at 1--18~m and 30 non-line-of-sight (NLOS) positions behind an industrial robot at 2.5--16.5~m across \frange{}. In LOS, the path-loss exponent ranged from 1.87 to 2.12, and the close-in fits agreed with the 3GPP InF-LOS model to within approximately 2~dB. However, shadow fading was only 1.4--1.7~dB, compared with the model’s 4.3~dB. In NLOS, the robot caused an excess loss that increased from 10.3 to 13.0~dB between 6.5 and 11.5~GHz, consistent with diffraction. Blocking the direct path raised the mean directional delay spread from approximately 5 to 25~ns, close to omnidirectional values. These results indicate that the 3GPP InF model captures mean LOS path loss well but overestimates the variability of beam-aligned FR3 links, and that blockage loss in FR3 is frequency dependent. The dataset and processing code are publicly available.
\end{abstract}


\section{Introduction}
\label{sec:intro}
Identifying spectrum that balances capacity and coverage is an important step towards sixth-generation (6G) mobile networks. The upper mid-band, commonly referred to as frequency range~3 (FR3) and spanning approximately 7--24~GHz, offers the potential for wider contiguous bandwidths than sub-6~GHz bands while retaining more favourable propagation conditions than millimetre-wave bands~\cite{bazzi2025upper, banerjee20266Gphy}. Interest in this range is reflected in international spectrum studies in preparation for the World Radiocommunication Conference 2027~\cite{itu2023resolution256}. In parallel, the Third Generation Partnership Project (3GPP) has undertaken dedicated work to validate and enhance channel models for this range~\cite{3gpp38901, tang2026rel19}. For private industrial networks supporting automation, collaborative robotics and extended-reality applications, reliable deployment requires channel models validated under representative factory conditions~\cite{reifert2024XR,rauf2026sionna}.

Industrial environments, such as factory halls, combine open production areas with large metallic structures, including machinery, shelving and enclosures. These structures give rise to strong specular reflections and can partially or completely obstruct direct propagation paths~\cite{umra2026RIS}. The indoor-factory (InF) model in 3GPP TR~38.901~\cite{3gpp38901} provides statistical descriptions of path loss, shadow fading and delay spread for omnidirectional channels. However, the scarcity of industrial FR3 measurements limits the empirical basis for assessing its underlying assumptions. In particular, further measurements are needed to assess whether the model's frequency-independent path-loss exponents and delay-spread statistics adequately describe industrial propagation across FR3.

Previous industrial channel measurements at sub-6~GHz and millimetre-wave frequencies have highlighted the importance of metallic reflectors and the differences between directional and omnidirectional channel statistics~\cite{schmieder2020industrial, schmieder2019directional}. More recent studies have examined frequencies within and adjacent to FR3. Factory measurements at 6.75 and 16.95~GHz across 12 transmitter--receiver locations provided directional and omnidirectional channel statistics, revealing frequency-dependent delay and angular spreads that differ from the 3GPP model~\cite{ying2025inf}. An omnidirectional measurement campaign covering 254 links at 10--12~GHz found that the sparse-clutter InF model described NLOS path loss better than the dense-clutter model~\cite{runtti2025industrial}. Outside industrial environments, continuous wideband measurements over 6--14~GHz demonstrated frequency-resolved channel characterisation in an urban microcell, although the campaign covered 11 receiver positions~\cite{abbasi2024umi}.

To complement industrial measurements at isolated carrier frequencies or over narrower bands, this paper combines contiguous frequency coverage with extensive spatial sampling under a consistent beam-aligned configuration. A measurement campaign using a vector network analyser (VNA) and boresight-aligned double-ridged horn antennas covers 135 line-of-sight (LOS) receiver positions and 30 non-line-of-sight (NLOS) positions behind an industrial robot in an industrial production hall of the Center for the Engineering of Smart Product-Service Systems (ZESS) at Ruhr University Bochum, Germany. Analysis in six contiguous 1-GHz sub-bands over 6--12~GHz characterises the close-in (CI) path-loss exponent~\cite{sun2016ci}, shadow fading, blockage-induced excess loss and root mean square (RMS) delay spread. The path-loss exponent and directional delay spread show no systematic frequency trend, while the 3GPP TR~38.901 InF-LOS model captures mean path loss well but overestimates shadow fading for the measured beam-aligned links. In NLOS, the excess loss caused by the robot increases with frequency, consistent with diffraction, and the directional delay spread rises to values close to omnidirectional ones. The dataset, including raw S-parameters, receiver positions and antenna gain data, is made publicly available together with the processing code\footnote{Dataset on GitHub: \url{https://github.com/FaiZanRaUf00/fr3-inf-zess-dataset}}. 
\section{Measurement Setup}
\label{sec:setup}

\subsection{Measurement Environment}
\label{sec:env}

\begin{figure*}[t]
    \centering
    \includegraphics[width=1\linewidth]{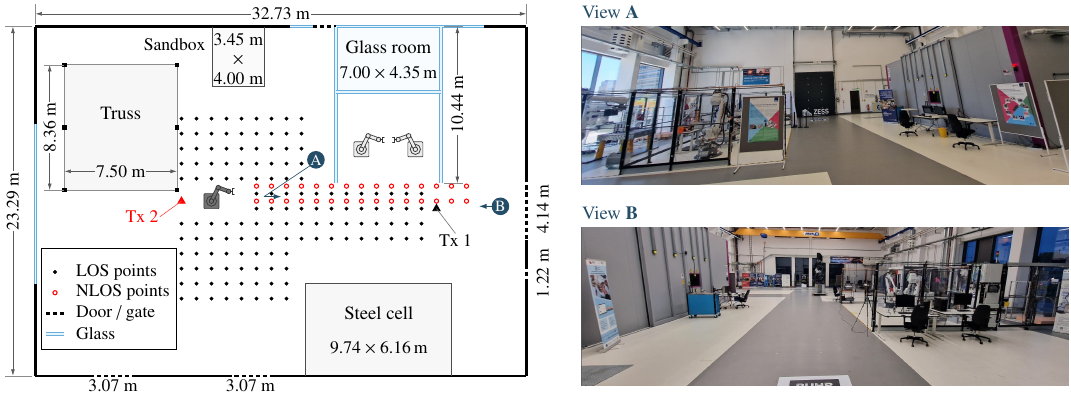}
    \caption{Measurement setup in the ZESS production hall, showing transmitter locations, LOS (black) and NLOS (red) measurement points, and photographic views A and B.}
    \label{fig:setup}
\end{figure*}

The measurements were carried out in the production hall of the ZESS, which has a floor area of $32.73 \times 23.29$~m$^2$ and a ceiling height of $10$~m. The hall contains an industrial robot cell enclosed by a safety fence of transparent panels in a metal frame, a steel truss structure, a sandbox test area, a glass-walled showroom and a closed steel enclosure, as shown in Fig.~\ref{fig:setup}. The measurement area is largely open, with the large metallic objects concentrated along the hall boundaries. Since both antennas are mounted well below the height of the surrounding objects and the clutter density is low, the environment corresponds most closely to the 3GPP InF sparse-clutter, low-base-station (InF-SL) sub-scenario~\cite{3gpp38901}.

For the LOS campaign, measurements were taken at 135 receiver positions with a visual line of sight to the transmitter, shown as black points in Fig.~\ref{fig:setup}. At some positions, the direct path passed through the transparent panels of the robot-cell fence. These panels introduced no significant additional loss, and all 135 positions were therefore classified as LOS. For the NLOS campaign, the transmitter was moved to a second location directly behind an industrial robot, which blocked the direct path to all 30 receiver positions. These positions, shown as red circles in Fig.~\ref{fig:setup}, were arranged in two parallel rows with 1-m spacing, spanning Tx--Rx distances of 2.5--16.5~m.
% \begin{figure}[t]
%     \centering
%     \includegraphics[width=\columnwidth]{Floortxrx.jpeg}\\[-1pt]
%     {\footnotesize (a)}\\[4pt]
%     \includegraphics[width=1\columnwidth]{zess_floorplan_v3.pdf}\\[-1pt]
%     {\footnotesize (b)}
%     \caption{Measurement setup in the ZESS production hall. (a) Horn antenna on a tripod, the VNA on a mobile cart and the robot cell with its transparent safety fence. (b) Floor plan with the transmitter (Tx) and the 135 receiver (Rx) positions on a 1-m grid.}
%     \label{fig:setup}
% \end{figure}

\subsection{Measurement System}
\label{sec:system}

The channel was measured in the frequency domain with a two-port Rohde~\&~Schwarz ZNA vector network analyser (VNA). The transfer function $S_{21}(f)$ was recorded from 6 to 17~GHz with 11\,001 points, i.e.\ a frequency step of 1~MHz, which corresponds to a maximum unambiguous excess delay of 1~$\mu$s (300~m path length). The intermediate frequency (IF) bandwidth was set to 1~kHz and the source power to 10~dBm. At each position, ten consecutive sweeps were averaged in the complex (vector) domain, which lowers the noise floor by 10~dB with respect to a single sweep.

The VNA was calibrated at the antenna connectors with a full two-port unknown-through, open, short, match (UOSM) calibration using a 2.92-mm calibration kit, so that the cables and adapters are removed from the measured response. The calibration was verified by a direct through measurement before each measurement session, and the calibrated instrument state was stored and recalled across measurement days.

Identical Schwarzbeck BBHA~9120~L double-ridged broadband horn antennas were used at the transmitter and receiver. This antenna covers 3--40~GHz, with a typical isotropic gain rising from about 9~dBi at 6~GHz to about 13~dBi at 16~GHz, a typical 3-dB beamwidth of $32^\circ$ (E-plane) and $36^\circ$ (H-plane), and a cross-polarisation discrimination above 20~dB~\cite{schwarzbeck9120l}. Both antennas were vertically polarised. The antenna gain was removed at every frequency point, as described in Section~\ref{sec:processing}. The main measurement parameters are summarised in Table~\ref{tab:meas_params}.

\begin{table}[t]
\centering
\caption{Measurement Parameters}
\label{tab:meas_params}
\renewcommand{\arraystretch}{1.1}\footnotesize
\begin{tabular}{|l|l|}
\hline
\textbf{Parameter} & \textbf{Value} \\
\hline
Equipment                    & R\&S ZNA VNA \\
Measured range               & 6--17~GHz \\ 
Analysed sub-bands           & 6--12~GHz (6 $\times$ 1~GHz) \\
Frequency points             & 11\,001 (1~MHz step) \\
IF bandwidth                 & 1~kHz \\
Averaging                    & 10 sweeps (vector) \\
Transmit power               & 10~dBm \\
Calibration                  & 2-port UOSM (2.92~mm) \\
Delay res. (sub-band)        & 1~ns \\
Maximum excess delay         & 1~$\mu$s \\
Tx / Rx antennas             & BBHA~9120~L horns \\
Antenna alignment            & Boresight to boresight \\
Tx / Rx height               & 1.5~m / 1.5~m \\
Rx positions                 & 135 (1-m grid), LOS / 30 (1-m grid), NLOS \\
Tx--Rx distance              & 1.0--18.0~m \\
\hline
\end{tabular}
\end{table}

\subsection{Measurement Procedure}
\label{sec:procedure}

The transmitter was placed at a fixed position at a height of 1.5~m. The 135 receiver positions were arranged on a regular grid with 1-m spacing in both directions at the same height, giving Tx--Rx distances between 1.0 and 18.0~m. At every position both horns were rotated towards each other, so that the link was aligned on boresight at both ends; the reported results therefore describe the boresight-aligned directional channel. Because both antennas are at the same height, the 3D Tx--Rx distance equals the horizontal distance, which was obtained from the grid geometry and a measured reference distance of 1~m to the nearest position.

At each position the averaging buffer was cleared after the antennas had been aligned, and the S-parameters were stored only after all ten sweeps had completed. During each sweep sequence of about 2~min the measurement area was kept free of moving people and objects, so that the channel can be considered static.

Both antennas operate in the radiating near field at the shortest distances: with the largest aperture dimension $D \approx 0.14$~m, the Fraunhofer distance $2D^2/\lambda$ ranges from about 0.8~m at 6~GHz to about 2.2~m at 16~GHz. All positions were retained in the model fits; the effect of the near field on the band-averaged path loss is discussed in Section~\ref{sec:res_pl}.

\section{Data Processing}
\label{sec:processing}

\subsection{Antenna De-Embedding and Sub-Band Path Loss}
\label{sec:pl_proc}

Since the calibration reference plane lies at the antenna connectors, the measured response $S_{21}(f)$ contains the propagation channel and the gains of both antennas. The antenna gains $G_\mathrm{t}(f)$ and $G_\mathrm{r}(f)$ were taken from the manufacturer's gain data~\cite{schwarzbeck9120l}, which are given in 100-MHz steps, and interpolated linearly to every measured frequency. The channel transfer function is then
\begin{equation}
    H(f) = \frac{S_{21}(f)}{\sqrt{G_\mathrm{t}(f)\,G_\mathrm{r}(f)}},
    \label{eq:deembed}
\end{equation}
with the gains in linear units. Removing the gains at every frequency point, rather than once per sub-band, matters because the horn gain rises by about $3.6$~dB across the analysed range.

The band is divided into $K$ = 6 non-overlapping sub-bands of 1~GHz width, each containing $N \approx 1000$ frequency points. For position $i$ and sub-band $k$ with centre frequency $f_k$, the path loss is computed from the mean received power,
\begin{equation}
    \mathrm{PL}_{i,k} = -10\log_{10}\!\left(\frac{1}{N}\sum_{n=1}^{N} \bigl|H_i(f_n)\bigr|^2\right) \quad \text{[dB]},
    \label{eq:pl}
\end{equation}
where the sum runs over the frequencies of sub-band $k$. Averaging over 1~GHz removes the small-scale fading, so that $\mathrm{PL}_{i,k}$ represents the local-area mean path loss. The NLOS measurements are processed in the same way, with their Tx--Rx distances taken from the NLOS transmitter position.

\subsection{Path-Loss Models}
\label{sec:pl_models}

For each sub-band and propagation condition, the close-in (CI) free-space reference distance model~\cite{sun2016ci} is fitted,
% \begin{equation}
%     \mathrm{PL}^\mathrm{CI}(f_k, d) = \mathrm{FSPL}(f_k, d_0) + 10\, n_k \log_{10}\!\left(\frac{d}{d_0}\right) + X_{\sigma,k},
%     \label{eq:ci}
% \end{equation}
\begin{equation}
    \begin{split}
        \mathrm{PL}^{\mathrm{CI}}(f_k,d)
        &= \mathrm{FSPL}(f_k,d_0) \\
        &\quad + 10\,n_k\log_{10}\!\left({d/d_0}\right)
        + X_{\sigma,k},
    \end{split}
    \label{eq:ci}
\end{equation}
with $d_0 = 1$~m and $\mathrm{FSPL}(f_k, d_0) = 20\log_{10}(4\pi f_k d_0/c)$. The path-loss exponent (PLE) $n_k$ is obtained by minimum mean-square-error fitting,
\begin{equation}
    n_k = \frac{\sum_i D_i A_{i,k}}{\sum_i D_i^2},
    \label{eq:ple}
\end{equation}
with $D_i = 10\log_{10}(d_i/d_0)$ and $A_{i,k} = \mathrm{PL}_{i,k} - \mathrm{FSPL}(f_k, d_0)$, and the shadow-fading standard deviation $\sigma_k$ is the standard deviation of the fit residuals. Because the CI model has a single free parameter, it remains stable over a limited distance range~\cite{sun2016ci}. As a cross-check, the floating-intercept model $\mathrm{PL} = \alpha_k + 10\beta_k\log_{10}(d)$ is also fitted; its slope $\beta_k$ is insensitive to a constant offset in the antenna gain. In NLOS, the slope $\beta_k$ is additionally compared with the 3GPP NLOS models, which are floating-intercept models as well. For NLOS, the excess loss caused by the blockage is evaluated as the difference between the measured path loss and the LOS CI fit at the same distance,
\begin{equation}
    \Delta\mathrm{PL}_{i,k} = \mathrm{PL}_{i,k} - \mathrm{FSPL}(f_k, d_0) - n_k^\mathrm{LOS} D_i,
    \label{eq:xl}
\end{equation}
where $n_k^\mathrm{LOS}$ is the CI exponent of the LOS positions in sub-band $k$, and $\Delta\mathrm{PL}_{i,k}$ is averaged over all NLOS positions $i$ in each sub-band.

\subsection{Delay Spread}
\label{sec:ds_proc}

The channel impulse response of each sub-band is obtained by an inverse discrete Fourier transform of $H(f)$ after applying a Hann window to suppress delay-domain sidelobes. The sub-band response is zero-padded by a factor of four, which interpolates the power delay profile (PDP) $P(\tau)$ but does not improve the delay resolution of 1~ns per sub-band. Multipath components weaker than 25~dB below the PDP peak are discarded, and the RMS delay spread is computed as
\begin{equation}
    \tau_\mathrm{rms} = \sqrt{\frac{\sum_j (\tau_j - \bar\tau)^2 P(\tau_j)}{\sum_j P(\tau_j)}}, \qquad
    \bar\tau = \frac{\sum_j \tau_j P(\tau_j)}{\sum_j P(\tau_j)}.
    \label{eq:ds}
\end{equation}
Since the received power is considerably lower in NLOS, multipath components less than 5~dB above the noise floor, estimated from the late part of the PDP, are also discarded, following~\cite{ying2025inf}.

\subsection{3GPP Reference Models}
\label{sec:3gpp}
\begin{figure*}[h!]
    \centering
    \includegraphics[width=1\textwidth, height=12cm]{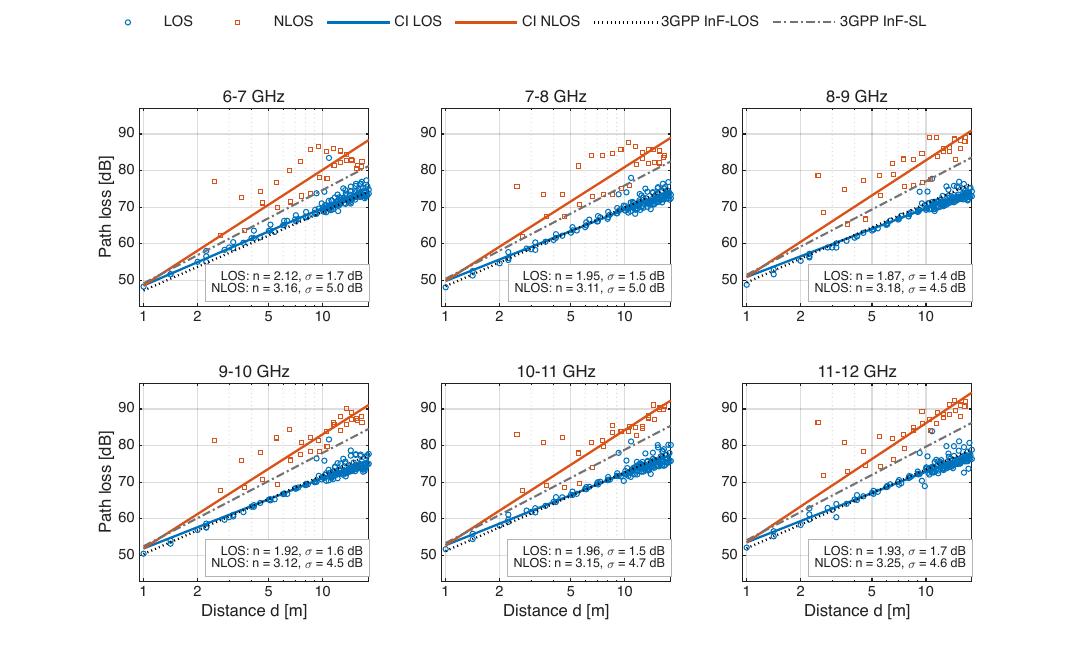}
    \caption{Measured path loss versus Tx--Rx distance in the \nbands{} 1-GHz sub-bands for the LOS positions and the NLOS positions behind the robot, with CI model fits ($d_0 = 1$~m) and the 3GPP InF-LOS and InF-SL models.}
    \label{fig:pl_subbands}
\end{figure*}

The results are compared with the 3GPP TR~38.901 InF model~\cite{3gpp38901}. With $d = d_\mathrm{3D}$ in m and $f_\mathrm{c}$ in GHz, the LOS path loss and the sparse-clutter (InF-SL) and dense-clutter (InF-DL) low-base-station NLOS models are
\begin{align}
    \mathrm{PL}_\mathrm{LOS} &= 31.84 + 21.50\log_{10} d + 19.00\log_{10} f_\mathrm{c}, \label{eq:3gpp_los}\\
    \mathrm{PL}_\mathrm{SL}  &= 33.00 + 25.50\log_{10} d + 20.00\log_{10} f_\mathrm{c}, \label{eq:3gpp_sl}\\
    \mathrm{PL}_\mathrm{DL}  &= 18.60 + 35.70\log_{10} d + 20.00\log_{10} f_\mathrm{c}, \label{eq:3gpp_dl}
\end{align}
with shadow-fading standard deviations of 4.3, 5.7 and 7.2~dB, respectively; the NLOS models are bounded below by the LOS model. The LOS model corresponds to a PLE of 2.15 at every frequency, and the InF-SL and InF-DL models to slopes of 2.55 and 3.57, respectively, which are compared with the measured NLOS slope $\beta_k$. These 3GPP statistics refer to omnidirectional antennas, whereas the measured values describe the boresight-aligned directional channel.
% needs \usepackage{multirow} in the preamble
\begin{table*}[!t]
\centering
\caption{Directional path-loss and delay-spread parameters per sub-band ($d_0=1 \mathrm{m}$), compared with other indoor-factory measurements and 3GPP}
\label{tab:params}
\footnotesize\setlength{\tabcolsep}{3.5pt}
\begin{tabular}{|l|c|c|c|c|ccc|cc|cc|}
\hline
\multirow{3}{*}{Campaign} & Band & Distance & Links & HPBW
 & \multicolumn{3}{c|}{LOS} & \multicolumn{2}{c|}{NLOS} & \multicolumn{2}{c|}{LOS DS [ns]} \\
\cline{6-12}
 & & & & & \multicolumn{2}{c}{CI} & FI & & $\sigma$ & & \\
 & [GHz] & [m] & (LOS/NLOS) & (Tx/Rx) & $n$ & $\sigma$ [dB] & $\beta$ & $n$ & [dB] & mean & 90\,\% \\
\hline
\multirow{6}{*}{\textbf{This work}}
 & 6--7   & \multirow{6}{*}{1--18} & \multirow{6}{*}{135/30} & \multirow{6}{*}{(34$^\circ$/34$^\circ$)} & 2.12 & 1.69 & 2.05 & 3.16 & 5.00 & 4.9 & 12.8 \\
 & 7--8   & & & & 1.95 & 1.45 & 1.97 & 3.11 & 4.99 & 4.9 & 16.0 \\
 & 8--9   & & & & 1.87 & 1.39 & 1.96 & 3.18 & 4.54 & 5.4 & 18.9 \\
 & 9--10  & & & & 1.92 & 1.59 & 1.98 & 3.12 & 4.49 & 5.2 & 15.2 \\
 & 10--11 & & & & 1.96 & 1.51 & 1.94 & 3.15 & 4.68 & 3.2 & 9.9 \\
 & 11--12 & & & & 1.93 & 1.74 & 1.91 & 3.25 & 4.60 & 5.7 & 18.7 \\
\hline
\multirow{2}{*}{~\cite{ying2025inf}}
 & 6.75  & \multirow{2}{*}{9--38} & \multirow{2}{*}{5/7} & (30$^\circ$/30$^\circ$) & 1.64 & 1.44 & -- & 2.92 & 7.69  & 11.2 & -- \\
 & 16.95 & & & (15$^\circ$/15$^\circ$) & 2.09 & 6.20 & -- & 3.81 & 10.42 & 13.7 & -- \\
\hline
\cite{ju2023142} & 142 & 5--87 & 30/48 & (8$^\circ$/8$^\circ$) & 2.10 & 1.80 & -- & 4.92 & 11.30 & 1.8 & -- \\
\hline
3GPP InF~\cite{3gpp38901} & 0.5--100 & 1--600 & -- & omni & 2.15 & 4.30 & 2.15 & 2.55--3.57 & 5.7--7.2 & -- & -- \\
\hline
\end{tabular}
\\[3pt]
\begin{minipage}{0.95\textwidth}
\end{minipage}
\end{table*}
\section{Results and Discussion}
\label{sec:results}

 \subsection{Path Loss}
\label{sec:res_pl}

Fig.~\ref{fig:pl_subbands} shows the measured path loss versus Tx--Rx distance in the six sub-bands, together with the CI fit and the 3GPP InF-LOS model. In all sub-bands the measurements follow a clear log-distance trend, and the six panels are nearly identical in shape. The only systematic difference between them is an offset: at a given distance, the path loss increases by about 5~dB from the 6--7 to the 11--12~GHz sub-band, which equals the increase of the free-space loss, $20\log_{10}(11.5/6.5) = 5.0$~dB. After the removal of the antenna gains, the frequency dependence of the LOS path loss is therefore fully described by the free-space term, and no additional frequency-dependent loss is observed between 6 and 12~GHz.

The CI fit and the 3GPP model agree within about 2~dB over the whole measured range. At short distances the CI fit lies slightly above the 3GPP model, because the 3GPP intercept is 1.4--1.7~dB below free space at 1~m, while at 13--17~m most measured points lie on or slightly below the 3GPP curve, because the measured distance dependence ($n = 1.87$--$2.12$) is slightly weaker than the 3GPP exponent of 2.15. The 1-m position agrees with free space within 2.1~dB in all sub-bands, which confirms the calibration and the antenna de-embedding; in the upper sub-bands it lies 1--2~dB below the CI fit, consistent with the near-field operation of the horns at this distance (Section~\ref{sec:procedure}). Since most positions of the grid lie between 9 and 17~m, the fits are dominated by the far positions, and the near-field positions have little influence on the fitted exponent.

Up to about 8~m the measured points lie within about $\pm 2$~dB of the fit. Beyond 10~m the scatter increases to about $\pm 3$~dB, and a few positions lie 5--10~dB above the fit in all sub-bands. Since the same positions are affected at every frequency, this excess is attributed to the geometry of these links, such as residual misalignment of the narrow beams or partial shadowing by the metal frame of the robot-cell fence, rather than to a frequency-dependent effect. 

The NLOS positions behind the robot lie well above the LOS points in all sub-bands. The difference is largest directly behind the robot, about 20~dB at 2.5~m, where the receiver lies deep in its shadow, and decreases to about 10~dB at the farthest positions, where diffraction around the robot and reflections from the walls and the steel enclosure fill the shadow region. As a consequence, the NLOS path loss increases only slowly with distance up to about 10~m and lies above the 3GPP InF-SL model at most positions, since the boresight-aligned horns point at the blocker and collect only part of the reflected power that an omnidirectional antenna would receive.
% Table II: either \input the file written by the script, or paste its contents here.

\subsection{Frequency Dependence of the Path-Loss Parameters}
\label{sec:res_freq}

\begin{figure}[h!]
    \centering
    \includegraphics[width=1\columnwidth]{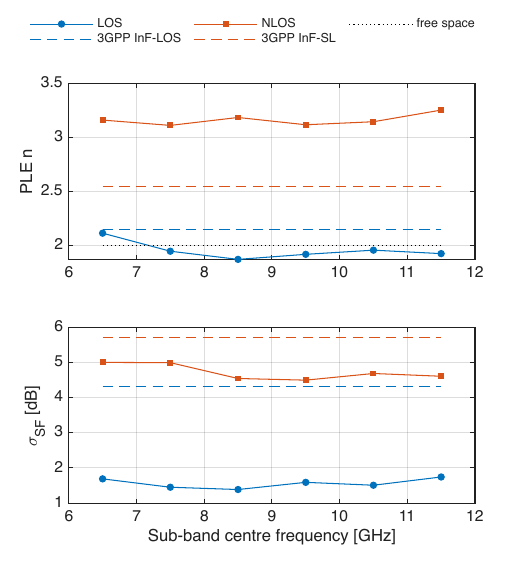}
    \caption{CI path-loss exponent (top) and shadow-fading standard deviation (bottom) versus sub-band centre frequency, compared with free space and the 3GPP InF model.}
    \label{fig:ple_freq}
\end{figure}

As shown in Fig.~\ref{fig:ple_freq}, the CI exponent stays between 1.87 and 2.12 without a frequency trend, supporting the frequency-independent 3GPP exponent of 2.15; the higher value at 6.5~GHz is attributed to the antenna-gain tolerance, since the offset-insensitive FI slope $\beta$ varies by only 0.14 (Table~\ref{tab:params}). The shadow fading of 1.4--1.7~dB is about three times lower than the 3GPP value of 4.3~dB, because the horns suppress multipath outside their beam. This agrees with the directional LOS measurements in~\cite{ying2025inf}, which report a shadow fading of 1.44~dB at 6.75~GHz with a similar beamwidth of $30^\circ$; their lower exponent of 1.64 is consistent with their longer links of 9--38~m, where more reflections fall into the beam. For 95\,\% coverage, 3GPP thus overestimates the fading margin of beam-aligned links by more than 4~dB.

In NLOS, the CI exponent of 3.11--3.25 also shows no frequency trend and lies between the 3GPP InF-SL and InF-DL slopes of 2.55 and 3.57, and the shadow fading of 4.5--5.0~dB is below both 3GPP values. The CI exponent is, however, inflated by the fixed free-space intercept at 1~m: the floating-intercept slope is only $\beta = 2.00$--$2.25$, i.e.\ the blockage mainly shifts the path loss by a large offset rather than steepening its distance dependence. This offset depends on frequency. The mean excess loss over the LOS fit in~\eqref{eq:xl} increases from 10.3~dB at 6.5~GHz to 13.0~dB at 11.5~GHz, by about 0.4~dB per GHz, which matches the increase of $10\log_{10}(11.5/6.5) = 2.5$~dB expected for diffraction around an obstacle. In contrast to the LOS path loss, the NLOS path loss in FR3 therefore contains a frequency-dependent component, which measurements at two frequencies cannot resolve.

\subsection{Delay Spread}
\label{sec:res_ds}

The mean RMS delay spread is 4.9--5.7~ns in all sub-bands except 10--11~GHz (3.2~ns), and the 90th percentile is 10--19~ns (Table~\ref{tab:params}), without a systematic frequency trend, consistent with the frequency-independent 3GPP InF delay spread. The distribution is strongly skewed: in more than half of the positions a single path dominates inside the antenna beams and the delay spread lies at the resolution limit, which an evaluation over the full 6--12~GHz bandwidth (0.17~ns resolution) confirms. Our mean values lie between those of the other directional factory campaigns, 11.2--13.7~ns at 6.75 and 16.95~GHz~\cite{ying2025inf} and 1.8~ns at 142~GHz with $8^\circ$ beams~\cite{ju2023142}; the larger values in~\cite{ying2025inf} are consistent with their angle-swept measurements and longer links of 9--38~m. All directional values are far below the omnidirectional 3GPP InF median for the measured hall. For beam-aligned FR3 links in such halls, the delay spread is therefore far below the cyclic prefix of NR numerologies, so that time dispersion is not a limiting factor.

In NLOS, the mean delay spread rises to 23.4--26.9~ns and the 90th percentile to 53--60~ns, again without a frequency trend. With the direct path blocked, no single path dominates, and the horns receive several delayed reflections inside their beam. The directional NLOS delay spread thus approaches omnidirectional values, such as the 29.7~ns measured in NLOS at 10--12~GHz in~\cite{runtti2025industrial}, i.e.\ the spatial filtering that keeps the LOS delay spread low is largely lost under blockage. Even in NLOS, however, the delay spread remains far below the cyclic prefix of NR numerologies.
\section{Conclusion}
\label{sec:conclusion}
This paper presented a wideband, sub-band-resolved channel measurement campaign in an industrial production hall in the FR3 upper mid-band. Using a VNA and horn antennas at 135 LOS and 30 NLOS positions, path loss, shadow fading, blockage-induced excess loss and RMS delay spread were extracted in six 1-GHz sub-bands over 6--12~GHz and compared with the 3GPP TR~38.901 InF model. In LOS, the PLE remains between 1.87 and 2.12, and the CI fits lie within about 2~dB of the 3GPP InF-LOS model, while the shadow fading of 1.4--1.7~dB is far below the 3GPP value of 4.3~dB. The robot caused an excess loss that increased from 10.3 to 13.0~dB between 6.5 and 11.5~GHz and raised the directional delay spread from about 5 to 25~ns. For beamformed FR3 links, the 3GPP InF model is thus accurate in the mean LOS path loss but conservative in its variability, while blockage loss must be treated as frequency dependent. Future work will consider different transmitter heights, as in the InF-SH and InF-DH sub-scenarios, MIMO measurements with the FR3 antenna array, and ray-tracing validation of the measured hall.

\bibliographystyle{IEEEtran}
\bibliography{ref}

\end{document}